\documentclass{article}
\usepackage{spconf,amsmath,graphicx}

\usepackage{amssymb}
\usepackage[ruled, vlined]{algorithm2e}
\usepackage{xcolor}
\usepackage{hyperref}
\usepackage{overpic}
\usepackage{subcaption}

\def\gtscale{0.12}
\def\hmscale{0.16}
\def\cbarscale{0.16}
\newcommand{\R}{\mathbb{R}}
\newcommand{\C}{\mathbb{C}}
\newcommand{\N}{\mathcal{N}}
\newcommand{\CN}{\mathcal{CN}}
\newcommand{\eps}{\epsilon}
\newcommand{\Id}[1]{\mathbf{I_{#1}}}

\newcommand{\x}{\mathbf{x}}
\newcommand{\y}{\mathbf{y}}
\newcommand{\A}{\mathbf{A}}

\newcommand{\bb}{\mathbf{b}}
\newcommand{\uu}{\mathbf{u}}
\newcommand{\rr}{\mathbf{r}}

\newcommand{\wl}{\mathbf{\Psi}}

\newcommand{\norm}[1]{\left\lVert#1\right\rVert}

\newcommand{\dvg}[2]{\text{div}_{#1}\left(#2\right)}

\newcommand{\tr}[1]{\text{tr}\left(#1\right)}
\newcommand{\diag}[1]{\text{diag}\left(#1\right)}

\newcommand{\Section}[1]{\vspace{-10pt}\section{#1}\vspace{-10pt}}
\newcommand{\Subsection}[1]{\vspace{-14pt}\subsection{#1}\vspace{-7pt}}

\title{SUREmap: Predicting Uncertainty in CNN-based Image Reconstructions using Stein's Unbiased Risk Estimate}

\name{Ruangrawee Kitichotkul, Christopher A. Metzler, Frank Ong, Gordon Wetzstein\thanks{R.K.~was supported by the Stanford Research Experience for Undergraduates (REU) program. C.M.~was supported by an appointment to the Intelligence Community Postdoctoral Research Fellowship Program at Stanford University administered by Oak Ridge Institute for Science and Education (ORISE) through an interagency agreement between the U.S. Department of Energy and the Office of the Director of National Intelligence (ODN). G.W.~was supported by an NSF CAREER Award (IIS 1553333), a Sloan Fellowship, and a PECASE by the ARL.}
\address{Department of Electrical Engineering at Stanford University\\
\texttt{\{rk22,cmetzler\}@stanford.edu}}}
\address{Stanford University\\
Department of Electrical Engineering\\
350 Jane Stanford Way, Stanford, CA}

\begin{document}
\setlength{\abovedisplayskip}{3pt}
\setlength{\belowdisplayskip}{3pt}
\setlength{\abovedisplayshortskip}{0pt}
\setlength{\belowdisplayshortskip}{0pt}

%\ninept
%
\maketitle
\begin{abstract}
Convolutional neural networks (CNN) have emerged as a powerful tool for solving computational imaging reconstruction problems. However, CNNs are generally difficult-to-understand black-boxes. Accordingly, it is challenging to know when they will work and, more importantly, when they will fail. This limitation is a major barrier to their use in safety-critical applications like medical imaging: Is that blob in the reconstruction an artifact or a tumor?

In this work we use Stein's unbiased risk estimate (SURE) to develop per-pixel confidence intervals, in the form of heatmaps, for compressive sensing reconstruction using the approximate message passing (AMP) framework with CNN-based denoisers. These heatmaps tell end-users how much to trust an image formed by a CNN, which could greatly improve the utility of CNNs in various computational imaging applications.
\end{abstract}
\Section{INTRODUCTION}
\label{sec:intro}

%What is CI
Computational imaging (CI) systems, like magnetic resonance imaging (MRI), can generally be described by the equation
\begin{equation}
    \y = \A\x + \eta,
\end{equation}
where $\y \in \C^m$ denotes the measurements, $\A \in \C^{m \times n}$ models the linear measurement operator/matrix, $\x \in \C^n$ is the vectorized latent image, and $\eta \in \C^m$ is additive noise. The goal of a computational imaging reconstruction algorithm is to reconstruct $\y$ from $\x$.

%What is CS. Some history of CS
When $m\ll n$ the reconstruction problem is underdetermined, and is known as compressive sensing (CS). CS reconstruction algorithms impose a prior, implicitly or explicitly, to form a reconstruction, $\hat{\x}$, of $\x$ from $\y$. While historically this prior was sparsity in some basis~\cite{candes2006robust}, the sparsity model has largely been superseded: Modern ``hand-designed'' methods achieve far better performance by imposing more elaborate priors, such as non-local self-similarity~\cite{NLRCS}. Meanwhile, learning-based methods, which impose priors with convolutional neural networks (CNNs), offer better performance still~\cite{ADMMnet}.

%How CNNs work
CNNs learn priors from vast quantities of training data, which they use to tune thousands to millions of parameters. In general,  it is unclear how each parameter contributes to the performance of the algorithm and it is difficult to know if and when a CNN-based method will successfully reconstruct an image.

%The problem with CNNs
Expected mean squared error (MSE), i.e.~risk, is the gold standard for evaluating a CS reconstruction algorithm. However, in general computing the risk requires access to the ground truth image -- which defeats the point of reconstruction in the first place. %A black-box approach is needed in order to evaluate the performance of these algorithms.

%How we overcome this problem
In this work, we demonstrate that when used in conjunction with the approximate message passing (AMP) framework~\cite{donoho2009amp}, which decouples the CS reconstruction problem into a series of additive white Gaussian noise (AWGN) denoising problems, one can accurately calculate the expected per-pixel MSE associated with CS reconstruction using Stein's unbiased risk estimate (SURE)~\cite{stein1981sure}. Consequently, we can generate heatmaps of low-pass filtered per-pixel MSE estimates \emph{without requiring access to the latent image}. We also apply this framework to the Variable Density AMP (VDAMP) algorithm~\cite{millard2020approximate}, an MRI reconstruction algorithm which decouples the problem into a series of additive colored Gaussian noise denoising problems. These uncertainty heatmaps could inform end-users about the reliability of image reconstructions and could also serve as supplementary information for an artifact-removal algorithm~\cite{guo2019toward} or to guide an adaptive sampling strategy~\cite{ji2008bayesian}.

\Section{RELATED WORK}
\label{sec:relwork}

Researchers have long sought to qualify the uncertainty associated CNN-based reconstructions. The importance of this problem was recently highlighted in~\cite{antun2019instabilities,gottschling2020troublesome}, where the authors showed how slight perturbations to a compressively sampled MRI signal can lead to vastly different, but still plausible looking, reconstructions.

If one assumes the latent image lies in the range of a generative network, one can use RIP-like conditions to guarantee recovery when the network is sufficiently expansive~\cite{bora2017compressed,hand2018global} or invertible~\cite{gilbert2017towards}. By looking at the distribution of an invertible network's latent variables, one can then estimate the uncertainty associated with a reconstruction~\cite{ardizzone2018analyzing}.

Alternatively, when dealing with probabilistic neural networks, as exemplified by variational autoencoders~\cite{kingma2019introduction}, one can sample from $p(\hat{\x}|\y)$, and thereby reason about the variance, but not the bias, associated with the reconstruction $\hat{\x}$~\cite{adler2019deep}. Similarly, bootstrap and jacknife resampling methods~\cite{tygert2018compressed} as well as a combination of variational dropout and input-dependent noise models~\cite{tanno2019uncertainty} can be used to estimate the variance of a reconstruction. One can even train a CNN  to identify motion artifacts~\cite{kustner2018automated}.

The majority of these method however can only characterize the variance associated with the reconstruction. They do not accurately predict the mean squared error, which is effected by bias as well.
\\

\begin{figure}[t]
	\centering
	\begin{subfigure}[b]{0.45\textwidth}
		\centering
		\begin{overpic}[width=.32\textwidth]{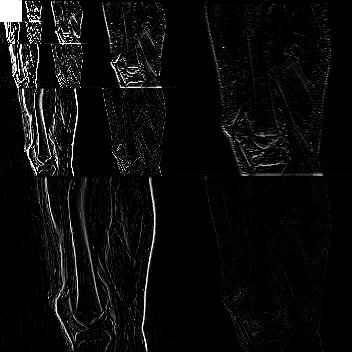}
%			\put (25,-14) {{\centering $\bm{w}=\bm{\Psi}\x$}}
			\put (15,-20){{\centering\begin{tabular}{@{}c@{}} True\\Coefficients \end{tabular}}}
		\end{overpic}
		\begin{overpic}[width=.32\textwidth]{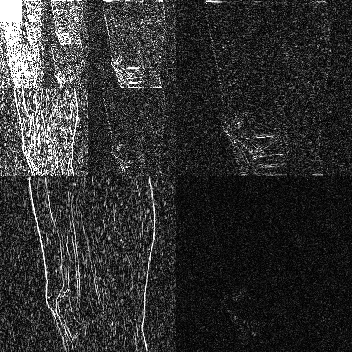}
%			\put (0,-14) {{\centering $\bm{r} = \bm{\Psi}\bm{F}^{-1}\bm{D}^{-1}\y$}}
			\put (15,-20){{\centering\begin{tabular}{@{}c@{}} Predicted\\Coefficients\end{tabular}}}
		\end{overpic}
		\begin{overpic}[width=.32\textwidth]{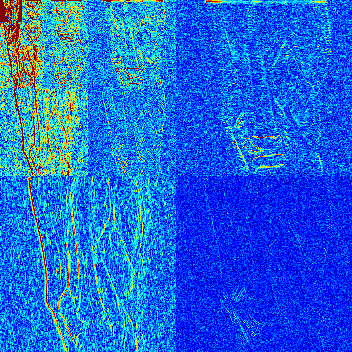}
%			\put (30,-14) {{\centering $|\bm{r}-\bm{w}|$}}
			\put (20,-20){{\centering\begin{tabular}{@{}c@{}} Absolute\\Difference \end{tabular}}}
		\end{overpic}
		\vspace{15 pt}
		\label{fig:ColoredNoise}
	\end{subfigure}
	\caption{{%\ninept
			{\bf DCLS effective noise.}  An illustration of the effective noise in the wavelet domain following a density compensated least squares reconstruction of a compressively sampled MRI image: The noise does not follow an i.i.d.~Gaussian distribution.}}
	\label{fig:effective_error}
	%	\vspace{-10 pt}
\end{figure}

Recently, Edupuganti et al.~predicted the per-pixel mean-squared error associated with reconstructed MRI images using SURE~\cite{edupuganti2020uncertainty}. However, in order to apply SURE, their method assumes that the difference between the true signal $\x$ and an initial estimate, formed with density compensated least squares (DCLS), follows a distribution that is both Gaussian and {\em white}.  As demonstrated in Figure~\ref{fig:effective_error}, the latter assumption does not hold in practice: The ``effective noise'', i.e.,~the difference between the estimate and the truth, demonstrates obvious structure when represented in the wavelet domain. These correlations invalidate the standard SURE approach, which applies only to i.i.d.~Gaussian noise.

\Section{Background}
\vspace{12pt}
\Subsection{Stein's Unbiased Risk Estimate}

SURE was first developed by its namesake several decades ago~\cite{stein1981sure}. Given a noisy signal $\y=\x+\eta$, where $\eta$ follows a Gaussian distribution with known covariance $\sigma^2\mathbf{I}$, SURE states that one can form an unbiased estimate of the mean squared error (MSE), $\frac{1}{n}\|\x-f(\y)\|^2$, via the expression
\begin{align}\label{eqn:SURE_loss}
 S{\big (}\y,f(\y){\big)}=\frac{1}{n}\|\y-f (\y)\|^2
-\sigma^2+\frac{2\sigma^2}{n}\text{div}_\y(f (\y)),
\end{align}
where $\text{div}_\y(f (\y))$ denotes its divergence, defined as
\begin{align}
\text{div}_\y(f (\y))=\sum_{n=1}^N \frac{\partial f_{n}(\y)}{\partial y_n}.
\end{align}

Since its introduction, SURE has been used extensively to tune algorithms. It is at the heart of the well-known SURE-shrink denoising algorithm~\cite{donoho1995adapting} and has been used extensively for tuning the parameters within various iterative reconstruction algorithms as well~\cite{mousavi2013parameterless,guo2015near,millard2020approximate,millard2020parameterfree}. SURE has also been combined with deep learning to train CNNs without ground truth data~\cite{metzler2018unsupervised,soltanayev2018suretrain,zhussip2019training} and been used to predict the error associated with a denoising algorithm's reconstruction~\cite{deledalle2012non}.

\begin{figure}[t]
	\centering
	\begin{subfigure}[b]{0.45\textwidth}
		\centering
		\begin{overpic}[width=.32\textwidth]{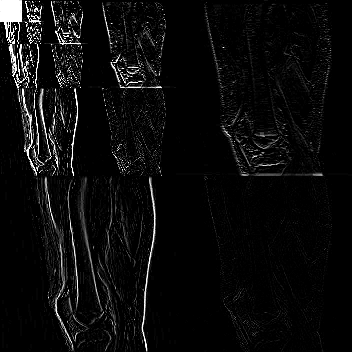}
			%			\put (25,-14) {{\centering $\bm{w}=\bm{\Psi}\x$}}
			\put (15,-20){{\centering\begin{tabular}{@{}c@{}} True\\Coefficients \end{tabular}}}
		\end{overpic}
		\begin{overpic}[width=.32\textwidth]{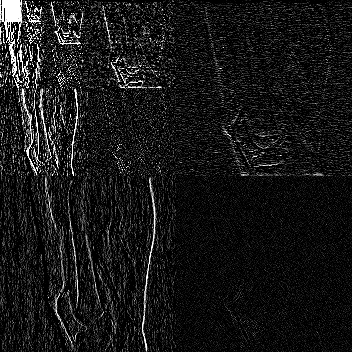}
			%			\put (0,-14) {{\centering $\bm{r} = \bm{\Psi}\bm{F}^{-1}\bm{D}^{-1}\y$}}
			\put (15,-20){{\centering\begin{tabular}{@{}c@{}} Predicted\\Coefficients\end{tabular}}}
		\end{overpic}
		\begin{overpic}[width=.32\textwidth]{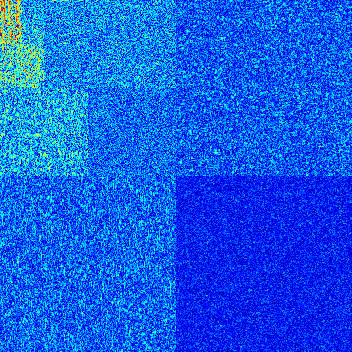}
			%			\put (30,-14) {{\centering $|\bm{r}-\bm{w}|$}}
			\put (20,-20){{\centering\begin{tabular}{@{}c@{}} Absolute\\Difference \end{tabular}}}
		\end{overpic}
		\vspace{15 pt}
	\end{subfigure}
	\caption{{%\ninept
			{\bf VDAMP effective noise.}  An illustration of the effective noise in the wavelet domain within an iteration of VDAMP while reconstructing a compressively sampled MRI image: The effective noise is approximately i.i.d.~within each wavelet subband.}}
	\label{fig:VDAMPEffectiveError}
	%	\vspace{-10 pt}
\end{figure}

\Subsection{Approximate Message Passing}
Approximate message passing (AMP), presented in Algorithm~\ref{algo:amp}, is a simple iterative algorithm for reconstructing a signal from i.i.d.~Gaussian measurements~\cite{donoho2009amp}, i.e.,~${\mathbf{A}_{i,j}\sim\mathcal{N}(0,1)}$ for all $i,j$. AMP resembles a projected gradient descent algorithm but comes with an additional term, $\frac{1}{m}\dvg{\mathbf{r}_t}{\x_{t + 1}}\mathbf{z}_t$, known as the Onsager correction. The Onsager correction ensures that at every iteration the effective noise, that is the difference between $\mathbf{r}_t$ and the ground truth signal $\x$, follows a white Gaussian distribution with variance $\hat{\sigma}^2$.

\begin{algorithm}[h]
    \SetKwInOut{Input}{Input}\SetKwInOut{Output}{Output}
    \SetAlgoLined
    \Input{Observation $\y \in \R^{m}$, Denoiser $f(\cdot)$, Measurement matrix $\A \in \R^{m \times n}$}
    \Output{Reconstructed image $\hat{\x}$}
    Initialize $\x_0 = \mathbf{0}_n, \ \mathbf{z}_0 = \y$\;
    \For{$t = 0, \dots, T - 1$}{
        $\mathbf{r}_t = \x_t + \A^T\mathbf{z}_t$ \\
        $\hat{\sigma}_t = \norm{\mathbf{z}_t}_2 / \sqrt{m}$ \\
        $\x_{t + 1} = f(\mathbf{r}_t; \hat{\sigma}_t)$ \\
        $\mathbf{z}_{t + 1} = \y - A\x_{t + 1} + \frac{1}{m}\dvg{\mathbf{r}_t}{\x_{t + 1}}\mathbf{z}_t$
    }
    \KwRet{$\x_T$}
\caption{AMP}
\label{algo:amp}
\end{algorithm}

Variable Density AMP (VDAMP) is a recent extension to AMP designed to solve the CS reconstruction problem when dealing with variable density sampled Fourier measurements~\cite{millard2020approximate}. Through multiscale updates in the wavelet domain, it ensures that the effective noise follows a colored Gaussian distribution with a {\em known covariance matrix}. This covariance matrix is diagonal when represented in the wavelet domain. Figure~\ref{fig:VDAMPEffectiveError} illustrates the empirical distribution of the effective noise associated with VDAMP.

While originally designed with simple, soft-thresholding based denoisers $f(\cdot)$, both AMP and VDAMP can be extended to incorporate more advanced denoisers, such as CNNs. The resulting Denoising-based AMP (D-AMP) and VDAMP (D-VDAMP) algorithms offer state-of-the-art performance when dealing with i.i.d.~Gaussian and variable density sampled Fourier measurements, respectively~\cite{metzler2016damp,metzler2017ldamp,DVDAMP}.

\Section{METHOD}
\label{sec:method}

In this work, we combined SURE with the denoising-based version of AMP and VDAMP to generate per-pixel mean-squared error estimates associated with reconstructions of compressively sampled images.

\Subsection{Uncertainty Quantification for D-AMP}

At each iteration, D-AMP solves a denoising problem described by
\begin{equation}
    \rr_t = \x + \eta_t; ~ \x, \rr_t \in \R^n;~ \eta_t \sim \N(0, \sigma_{\eta}^2 \Id{n}), \label{eq:amp-se}
\end{equation}
where $\eta_t$ is the noise at the $t$-th iteration, and $\rr_t$ is the corresponding noisy image. 
The final estimate formed by $T$ iterations of AMP is $ \hat{\x}=f(\rr_T)$.
Because this is the output of a simple AWGN denoising problem, SURE can be used to estimate the mean squared error associated with this reconstruction.

When a closed form expression for $\text{div}_\y(f_\theta (\rr_T))$ is not available, it can be estimated with the following Monte-Carlo estimate~\cite{ramani2008mcsure}
\begin{equation}
    \dvg{\rr_T}{\hat{\x}} \approx \frac{1}{K}\sum_{k=1}^K \frac{1}{\eps}\bb_k^T \left(f(\rr_T + \eps\bb_k) - f(\rr_T)\right),    \label{eq:mcsure}
\end{equation}
where $\bb_k \sim \N(0, \Id{n})$ and $\eps \in \R$ is a small number, chosen to be $\frac{\max(\rr_T)}{1000}$ in this work, and $K$ is the number of Monte-Carlo samples used in the approximation. 
To generate a per-pixel SURE heatmap, we compute SURE for overlapping patches of the reconstruction and average the result.

\Subsection{Uncertainty Quantification for D-VDAMP}

At each iteration, D-VDAMP solves a denoising problem described by
\begin{equation}
\rr_t = \x + \eta_t;~  \x, \rr_t \in \R^n;~ \eta_t \sim \CN(0, \mathbf{\Psi}^t\diag{\tau_t}\mathbf{\Psi}),   \label{eq:vdamp-se}
\end{equation}
where $\CN(0, \mathbf{\Psi}^t\diag{\tau_t}\mathbf{\Psi})$ denotes a circular Gaussian distribution with independent real and imaginary parts, each of which has mean $0$ and the covariance matrix $\frac{1}{2}\mathbf{\Psi}^t\diag{\tau_t}\mathbf{\Psi}$, and $\mathbf{\Psi}$ denotes the wavelet transform matrix. (We use a four-level 2-D Haar transform throughout this paper.) 
As before, the final estimate associated with the denoising-based version of VDAMP is $\hat{\x}=f(\rr_T)$.\footnote{The original VDAMP work, which was based on soft wavelet thresholding, included an additional gradient step after denoising $\rr_T$~\cite{millard2020approximate}. In~\cite{DVDAMP}, the authors found this term hurts the algorithm's performance when dealing with more advanced denoising algorithms, and so we do not adopt it here.}

As demonstrated in the Generalized SURE work~\cite{eldar2009gsure}, an unbiased risk estimate for removing colored Gaussian noise $\eta_T \sim \N(0, \Sigma)$ is
\begin{equation}
    S(\hat{\x}, \rr_T) = \frac{1}{n}\norm{\hat{\x} - \rr_T}^2 + \frac{2}{n}\dvg{\uu}{\hat{\x}} - \tr{\Sigma}, \label{eq:sure2}
\end{equation}
where $\uu = \Sigma^{-1}\rr_T$.

We can extend this estimate to the complex case by noting that $
\|\x-\hat{\x}\|^2=\|\mathcal{R}(\x-\hat{\x})\|^2+\|\mathcal{I}(\x-\hat{\x})\|^2.
$
We next note that with $\Sigma=\mathbf{\Psi}^t\diag{\tau_t}\mathbf{\Psi}$, the similarity invariance of the trace function implies $\tr{\Sigma}=\sum_{i = 1}^n \tau_T^{(i)}$. We are then left with
\begin{equation}
\begin{split}
S(\hat{\x}, \rr_T) &= \norm{\hat{\x} - \rr_T}^2 - \sum_{i = 1}^n \tau_T^{(i)} \\
&+ \frac{2}{n}\left(\dvg{\Re(\uu)}{\Re(\hat{\x})} + \dvg{\Im(\uu)}{\Im(\hat{\x})}\right),
\end{split} \label{eq:sure3}
\end{equation}
where $\uu =  \wl\diag{\frac{1}{2}\tau_t}^{-1}\wl^t \rr_T$.

To estimate the divergence, we let $\widetilde{f}(\uu) = f(\Sigma\uu)$. Now we have
\begin{align*}
\dvg{\uu}{f(\rr_T)}  = \dvg{\uu}{\widetilde{f}(\uu)},% \\
%    \dvg{\uu}{f(\y_T)} = \tr{\frac{\partial f(\Sigma\uu)}{\partial \uu}} = \dvg{\uu}{\widetilde{f}(\uu)},% \\
%    &= \tr{\frac{\partial \widetilde{f}(\uu)}{\partial \uu}} = \dvg{\uu}{\widetilde{f}(\uu)}.
\end{align*}
and can use the Monte-Carlo approximation \eqref{eq:mcsure} to obtain
\begin{equation}
    \dvg{\uu}{f(\rr_T)} \approx \frac{1}{K}\sum_{k=1}^K \frac{1}{\eps}\bb_k^T(f(\rr_T + \eps\Sigma\bb_k) - f(\rr_T)) \label{eq:mcsure2},
\end{equation}
which we apply independently to both the real and imaginary parts of $f(\rr_T)$.

As before, we generate per-pixel SURE heatmaps by averaging the overlapping estimated risks of square patches.

\Section{EXPERIMENT}
\label{sec:experiment}
\vspace{12pt}

\begin{figure*}[ht]
    \flushright
	\begin{overpic}[width=\gtscale\textwidth]{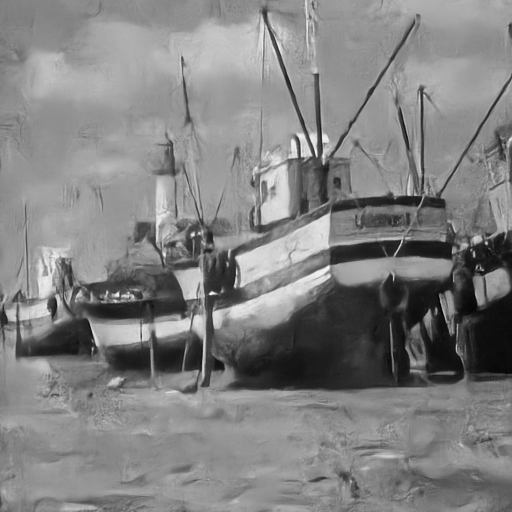}
        \put (-70, 40) {D-AMP}
        \put (0, 109) {Reconstruction}
	\end{overpic}
	\begin{overpic}[width=\hmscale\textwidth]{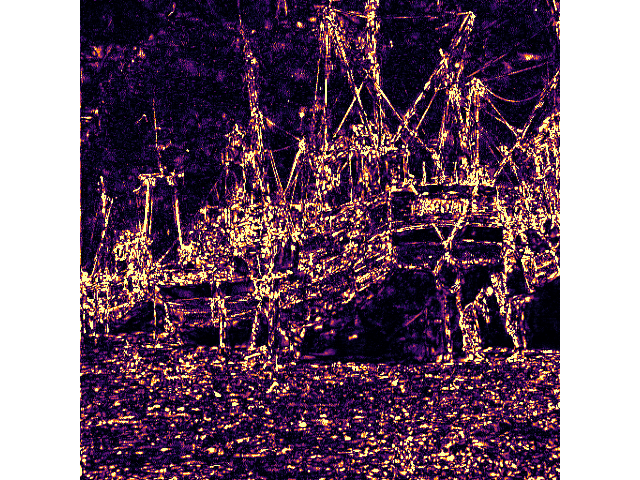}
        \put (37, 82) {MSE}
	\end{overpic}
    \begin{overpic}[width=\cbarscale\textwidth]{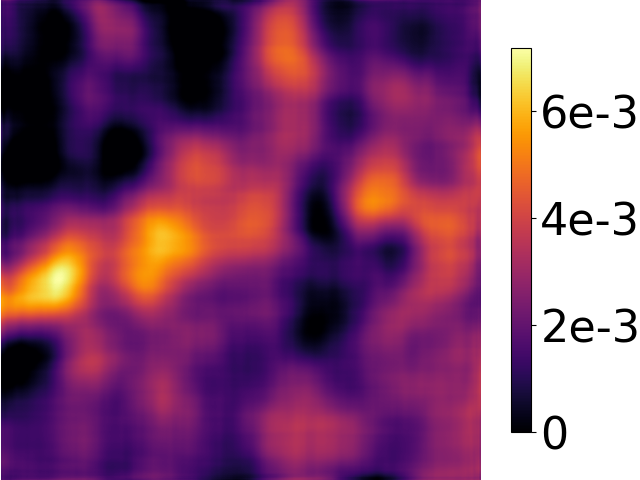}
        \put (21, 82) {SURE}
	\end{overpic}
    \begin{overpic}[width=\gtscale\textwidth]{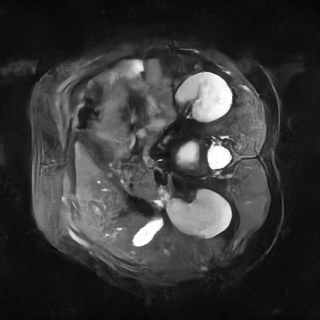}
        \put (0, 110) {Reconstruction}
	\end{overpic}
	\begin{overpic}[width=\hmscale\textwidth]{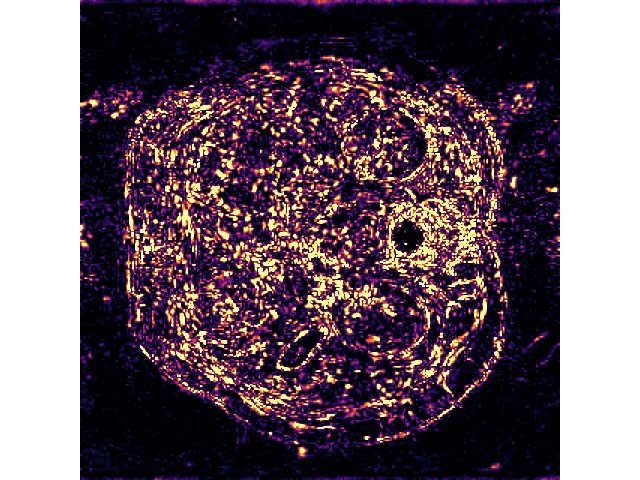}
        \put (37, 82) {MSE}
	\end{overpic}
    \begin{overpic}[width=\cbarscale\textwidth]{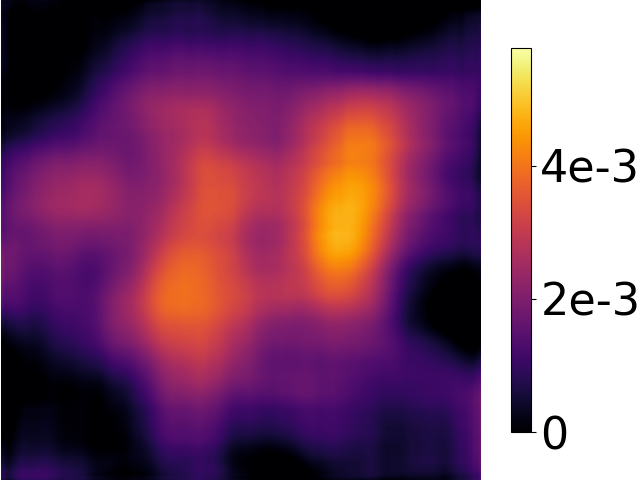}
        \put (21, 82) {SURE}
	\end{overpic}
    \begin{overpic}[width=\gtscale\textwidth]{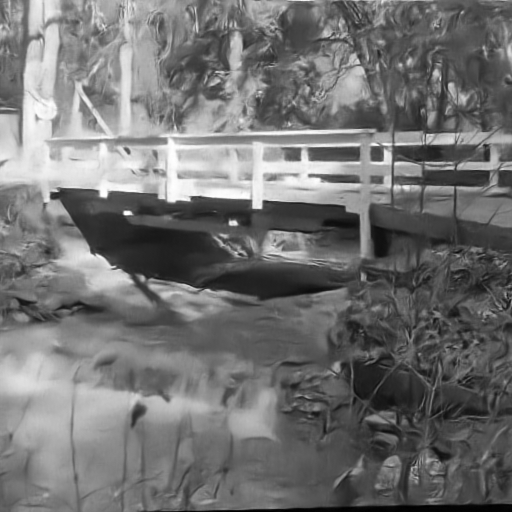}
        \put (-95, 40) {D-VDAMP}
	\end{overpic}
	\begin{overpic}[width=\hmscale\textwidth]{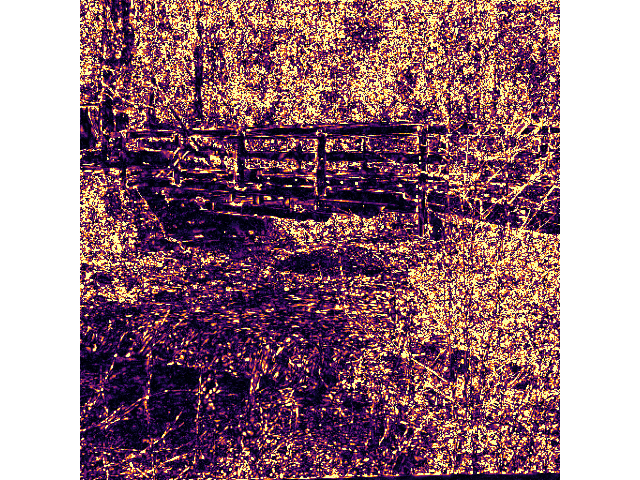}
	\end{overpic}
    \begin{overpic}[width=\cbarscale\textwidth]{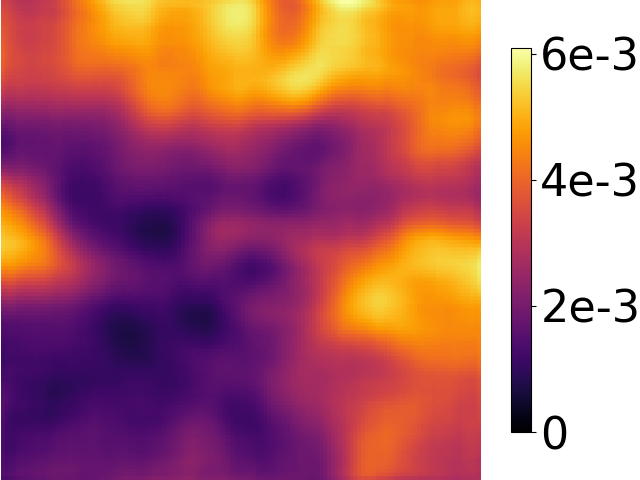}
	\end{overpic}
    \begin{overpic}[width=\gtscale\textwidth]{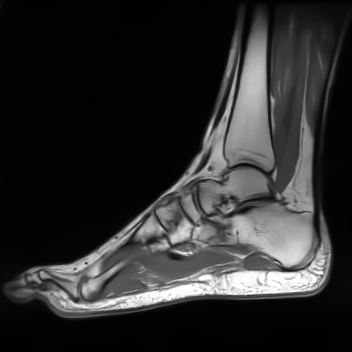}
	\end{overpic}
	\begin{overpic}[width=\hmscale\textwidth]{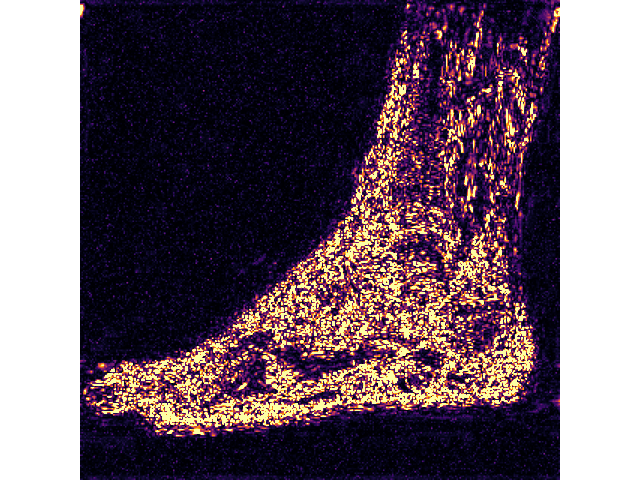}
	\end{overpic}
    \begin{overpic}[width=\cbarscale\textwidth]{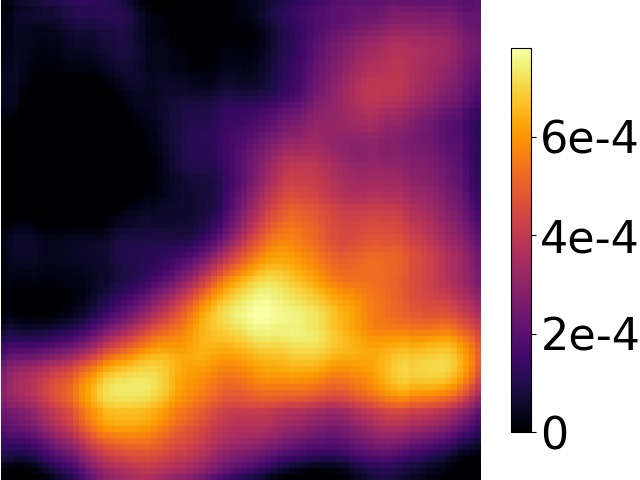}
	\end{overpic}
    \caption{{\bf SURE heatmaps and MSE heatmaps} of CS reconstructions with D-AMP and D-VDAMP along with the reconstructed images. For all heatmaps in this figure, the patch size is 48x48 pixels, and the number of Monte-Carlo samples for divergence estimation is 2. MRI images are from \url{MRIdata.org}.} 
    \label{fig:heatmap}
\end{figure*}

\vspace{12pt}
\Subsection{Setting}

We test our SURE heatmap generation method with CS reconstructions using D-AMP (Gaussian measurement matrices) and D-VDAMP (subsampled Fourier measurement matrices). For D-AMP, the sampling rate, $m/n$, is 5\% and the SNRs are 23dB and 18dB for the natural image and the MR image, respectively.  
For D-VDAMP, the sampling rate is 25\% and the SNR is 20dB. The Fourier coefficients were selected using polynomial variable density sampling \cite{lustig2007sparse}.
Both D-AMP and D-VDAMP used a collection of DnCNN~\cite{zhang2017dncnn} denoisers trained for multiple noise levels from $\sigma = 0$ to $\sigma = 300/255$. The natural images were $512\times512$ while the MR images were $320\times 320$.

\begin{figure}[t]
	\centering
	\begin{subfigure}{0.48\textwidth}
		\centering
		\begin{overpic}[width=.7\textwidth]{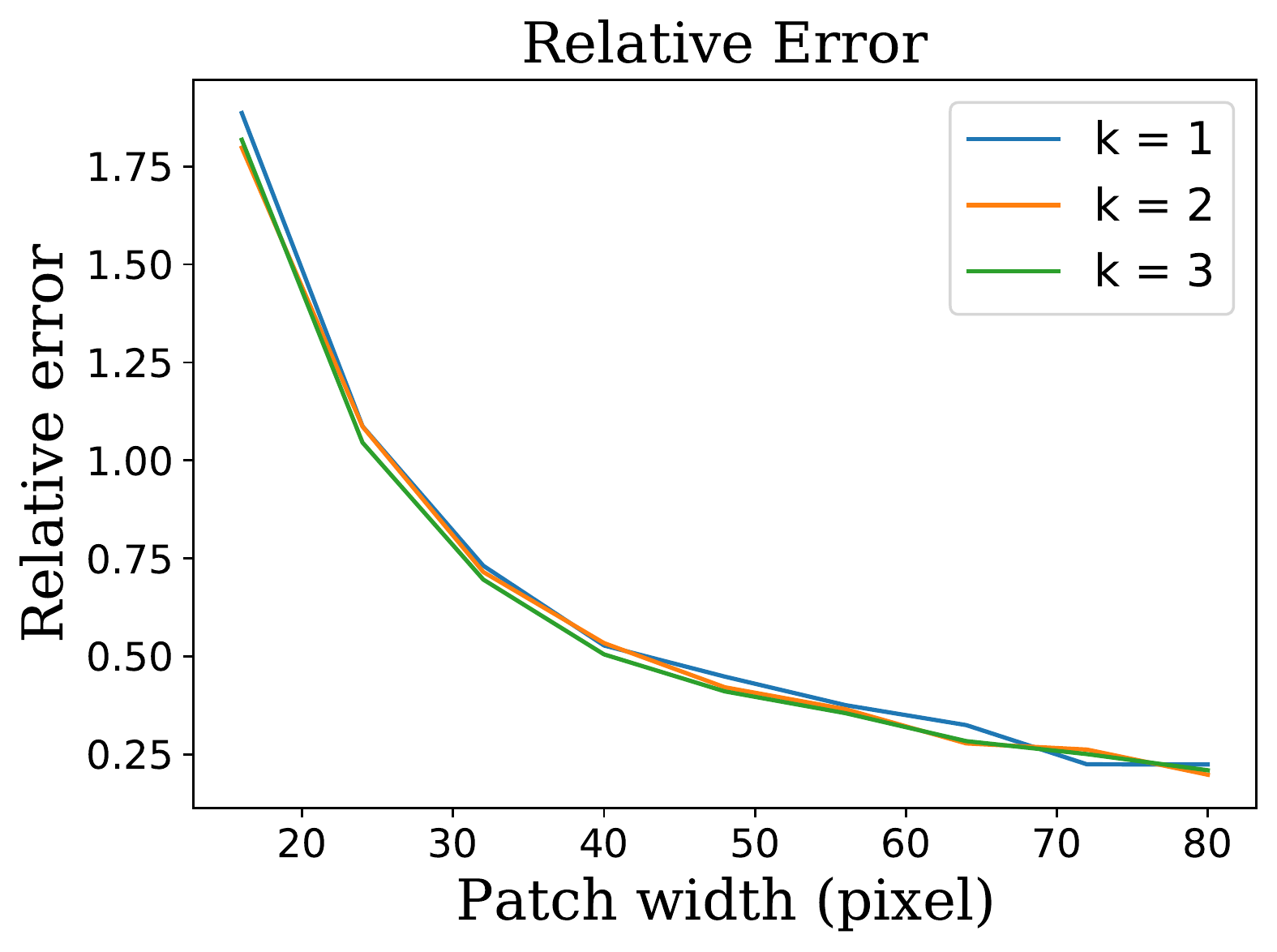}
			\put (40, 35) {{$|\text{MSE} - \text{SURE}|/\text{MSE}$}}
		\end{overpic}

	\end{subfigure}
	\caption{{{\bf Normalized absolute difference} between the SURE heatmap and the patch-average (effectively low-pass filtered) MSE heatmap, which is generated by averaging overlapping patches of MSEs in the same fashion as the SURE heatmap generation. Data is for a CS reconstruction using D-AMP.}}
	\label{fig:error}
\end{figure}

\Subsection{Accuracy-resolution trade-off}

\begin{figure}[ht]
	\centering
	\begin{subfigure}{0.48\textwidth}
		\centering
		\begin{overpic}[width=.32\textwidth]{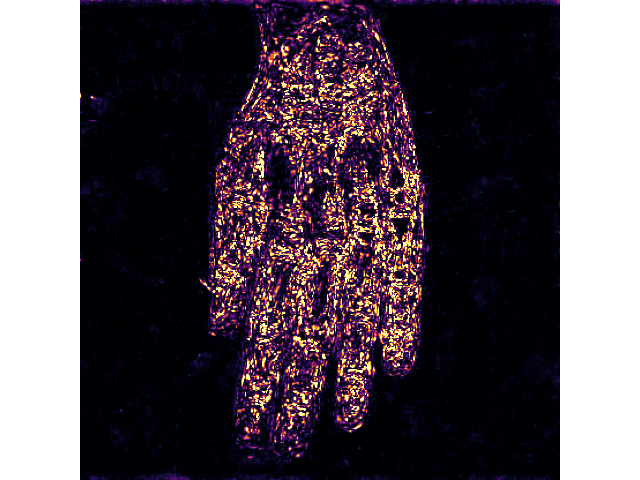}
			\put (35, 80) {MSE}
		\end{overpic}
		\begin{overpic}[width=.32\textwidth]{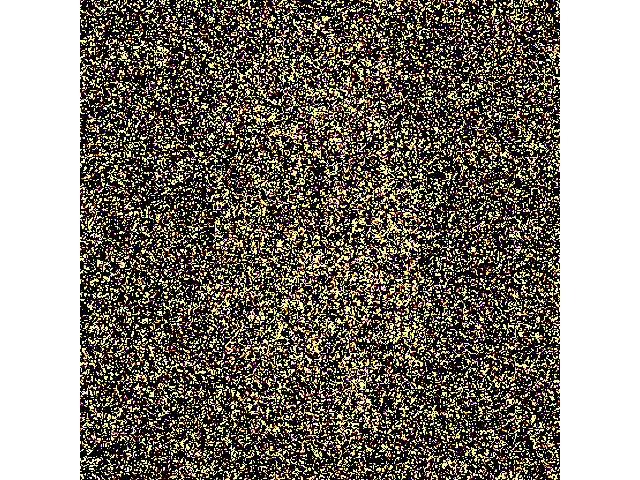}
			\put (10, 80) {Patch width = 1}
		\end{overpic}
		\begin{overpic}[width=.32\textwidth]{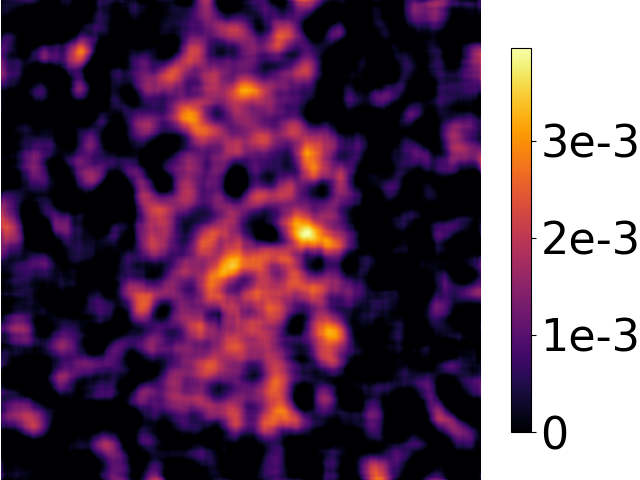}
			\put (-5, 80) {Patch width = 16}
		\end{overpic}
	\end{subfigure}
	\caption{{\bf SURE heatmaps with small patch sizes.} The left heatmap is the MSE. The middle and the right heatmaps are SURE heatmaps of a CS reconstruction with D-AMP generated by using patch widths of 1 pixel and 16 pixels respectively. The number of Monte-Carlo samples, $K$, is 3 for both heatmaps.}
	\label{fig:smallpatch}
\end{figure}

We first investigate the accuracy of the MSE estimate as a function of patchsize.
Figure~\ref{fig:error} compares the average difference squared between the SURE estimate and the true MSE of the image as one increases the patch sizes used in the SURE estimates. We observe that, due primarily to the reduced variance of the data fidelity term ($\norm{\hat{\x} - \rr_T}^2$), the SURE heatmaps become more accurate as the patch size increases. Increasing the number of Monte-Carlo samples, $K$, has only a slight effect on the accuracy of the estimate. 
Figure~\ref{fig:smallpatch} compares the heatmaps formed with various patch sizes. While smaller patch-sizes are higher resolution, larger patch sizes result in more accurate MSE estimates. We found $48\times 48$ patches provided a nice trade-off between resolution and accuracy.

\Subsection{Results}
Figure~\ref{fig:heatmap} generates the SURE heatmaps for D-AMP and D-VDAMP reconstructions using a patch size of $48\times48$ pixels. While somewhat low resolution, the shapes and magnitudes of the heatmaps closely follow the true pixelwise MSEs.
These heatmaps, which {\em do not require the ground truth}, could prove valuable for medical diagnosis and other safety-critical applications.

\Section{References}
\renewcommand{\section}[1]{}

{
	\ninept
\bibliographystyle{IEEEbib}
\bibliography{refs}
}
\end{document}